\documentclass{optica-article}

\journal{opticajournal} 

\articletype{Research Article}

\usepackage{lineno}

\begin{document}

\title{Quasi-Random Frequency Sampling for Optical Turbulence Simulations}

\author{Amokrane Berdja\authormark{1,2,*} Massinissa Hadjara\authormark{3,4}, Marcel Carbillet\authormark{5}, Rafael L. Bernardi\authormark{6}, and Romain G. Petrov\authormark{5}}

\address{\authormark{1}Centro de Astronom\'{i}a (CITEVA), Universidad de Antofagasta, Avenida Angamos 601, Antofagasta 1270300, Chile\\
\authormark{2}DICTUC S.A., Av. Vicu\~{n}a Mackenna 4860, 7820436 Santiago, Chile\\
\authormark{3}Nanjing Institute of Astronomical Optics \& Technology, National Astronomical Observatories, Chinese Academy of Sciences, Nanjing 210042, China\\
\authormark{4}Chinese High Angular Resolution Southern Astronomical Laboratory (CHARSAL)/Astro-Photonics Laboratory/Space and Planetary Exploration Laboratory (SPEL), Dept. of Electrical Engineering, FCFM, Universidad de Chile, Av. Tupper 2007, Santiago, Chile\\
\authormark{5}Laboratoire Lagrange (Universit\'e C\^ote d'Azur, Observatoire de la C\^ote d'Azur, CNRS), B\^at.\,Fizeau, Parc Valrose, 06100 Nice, France\\
\authormark{6}Department of Electrical Engineering, Pontificia Universidad Cat\'{o}lica de Chile, Vicu\~{n}a Mackenna 4860, 7820436 Santiago, Chile}

\email{\authormark{*}amokrane.berdja@uantof.cl} 


\begin{abstract*} 
Optical turbulence modelling and simulation are crucial for developing astronomical ground-based instruments, laser communication, laser metrology, or any application where light propagates through a turbulent medium. In the context of spectrum-based optical turbulence Monte-Carlo simulations, we present an alternative approach to the methods based on the Fast Fourier Transform (FFT) using a quasi-random frequency sampling heuristic. This approach provides complete control over the spectral information expressed in the simulated measurable, without the drawbacks encountered with FFT-based methods such as high-frequency aliasing, low-frequency under-sampling, and static sampling statistics. The methods's heuristics, implementation, and an application example from the study of differential piston fluctuations are discussed.
\end{abstract*}


\section{Introduction}
\label{sec:introduction}

Optical turbulence numerical simulation has become an essential tool in Astronomy, laser communication, and metrology.
It is used to model how ground-based instruments respond to optical turbulence in astronomy. For example, it is central to the modelling of Adaptive Optics (AO) performance and responsiveness~\cite{Carbillet2005, Emiel2018, Townson:19}. It can also be used to test the effectiveness of any instrument that estimates and monitors optical turbulence. 
Imaging applications, such as in AO, rely on simulating phase fluctuations, including scintillation, at a telescope's pupil plane after propagation through the atmosphere.
Several simulation methods that have been proposed, including the use of Zernike bases~\cite{Roddier1990, Whiteley:98, Chimitt2023}, but the most commonly used method is the spectrum-based Monte-Carlo simulation. It relies on a model of the 2D spatial power spectrum density of phase fluctuations (Kolmogorov, von K\'{a}rm\'{a}n, Greenwood-Tarazano, "exponential", etc.), depending on how the effect of the outer scale of optical turbulence is modelled~\cite{Lukin2021}. The implementation typically involves using the 2D Fast Fourier Transform (FFT) algorithm~\cite{Bracewell2000}. 

It has long been recognised that FFT-based simulations may suffer from certain drawbacks in practice. This is mainly due to limited computational resources, which can result in compromises in the sampling of spectra. The most well-documented issue is the under-sampling of low frequencies in the case of phase fluctuation simulations. Many approaches involving the addition of sub-harmonics have been applied to address this issue.
Another drawback is that the spectral under-sampling is static. This means that the statistical properties of the simulated quantity correspond to the averaged sampled power spectrum density rather than the theoretical average power spectrum density~\cite{Charnotskii:20}. 

In addition to these drawbacks there is also an issue with many applications, such as imaging through a circular aperture, where a significant portion of the simulated phase fluctuations goes unused. This waste is even more substantial in applications like stellar interferometry simulations when phase-screens are used~\cite{Porro:00}.

This paper proposes a solution for spectrum-based simulations of phase and phase-related optical turbulence quantities. The solution relies on a quasi-random sampling of the frequency space. This approach may solve many of the inconveniences encountered with conventional FFT-based simulations when used properly.

The following section introduces spectral methods and FFT-based simulations (refer to Section~\ref{sec:SpectrumSimul}). It also addresses issues that arise in practice when using the FFT-based approach. We then discuss how our approach deals with these issues (refer to Section~\ref{sec:QuasiRandom}). To demonstrate the statistical validity of the method, we apply it to simulate the differential piston fluctuations (refer to Section~\ref{sec:Application}). This serves as a test for the statistical properties of the simulated piston, as the structure function of these fluctuations can be predicted analytically.

\section{Spectrum-based and FFT-based simulations of optical turbulence}
\label{sec:SpectrumSimul}

\subsection{Spectrum-based simulation}
\label{subsec:spectrumbasedsimul}

The primary objective of of optical turbulence Monte-Carlo simulations is to generate pseudo-random fields of quantities such as phase fluctuations to replicate real measurements. The statistical properties of these fields are determined by a given model.

The primary statistical properties of interest are second order statistics such as the variance, the covariance, or the structure function. These statistical properties are directly related to the power spectrum density of the quantity being simulated as per the Wiener–Khinchin theorem. Therefore, a fluctuating quantity $\chi$ with an average power spectrum density $W_{\chi}$, which corresponds to a certain covariance/structure function, can be obtained from a randomised spectrum whose squared modulus averages to $W_{\chi}$. 
The following sections consider a 2D spatial field, which is commonly used in optical turbulence simulations. However, the concepts discussed can be generalised to temporal, 3D spatial, or spatial-temporal fields and power spectrum densities. In this context, we are discussing the simulation of a quantity $\chi(\vec{r})$, where $\vec{r}$ is a 2D spatial position vector, and its corresponding theoretical spatial power spectrum density $W_{\chi}(\vec{f})$, where $\vec{f}$ is a 2D spatial frequency vector. In this case, each frequency $\vec{f}$ contributes a random harmonic to the field $\chi(\vec{r})$.

Each random harmonic has a random amplitude and a random phase. As there are no constraints on the phase of these random harmonics, it is appropriate to assign them a uniform random distribution between $0$ and $2\pi$. In this case, it can be stated that the phase as a function of spatial frequencies is $2\pi R_{u}(\vec{f})$, where $R_{u} \sim \mathcal{U}(0,1)$ is a random (or pseudo-random) number with a uniform distribution between $0$ and $1$. 

Similarly, given that the only constraint on the random harmonic is to have a finite variance, it is reasonable to postulate as per the maximum entropy principle that the amplitude is normally distributed with a variance equal to $W_{\chi}(\vec{f})$. Therefore, we can represent the amplitude of the random harmonic as $R_{n}(\vec{f})\sqrt{W_{\chi}(\vec{f})}$, where $R_{n} \sim \mathcal{N}(0,1)$ represents a random number with a normal distribution and a variance of $1$.
This assumption may need to be updated if there are additional constrains on the spectrum from theory or observation.

With these considerations in mind, we can write a general encapsulation of Monte Carlo spectral density-based simulation methods as:

\begin{equation}
\label{eq:1}
\chi(\vec{r}) = \mathcal{F}^{-1}\left(R_{n}(\vec{f})\sqrt{W_{\chi}(\vec{f})} \exp{\left( 2\pi i R_{u}(\vec{f}) \right)} \right).
\end{equation}

\subsection{FFT-based implementation}

Most implementations of the spectral-density approach rely on the numerical efficiency of the FFT algorithm~\cite{Bracewell2000}. The 2D FFT implementations, especially for phase fluctuations, are widely used in the simulation of isoplanatic and anisoplanatic imaging applications~\cite{Bos2012, Hardie2022}, as well as in AO in astronomy, to name just a few examples. The use of 2D phase fluctuation matrices is very convenient for introducing such effects as Fresnel diffraction~\cite{Schmidt:22} within the Fourier optics framework~\cite{Goodman2005} to simulate various light propagation through turbulence phenomena such as scintillation~\cite{Johnston:00}.

Based on the above formulation, most 2D FFT implementations can be represented as~\cite{Schmidt2010}:

\begin{equation}
\label{eq:2}
\chi(\vec{r}) = \mathcal{FFT}^{-1}\left( (R_{n}(\vec{f}) + i R_{n}(\vec{f}))  \sqrt{W_{\chi}(\vec{f})} \right).
\end{equation}

\noindent It is worth noting here that the two $R_{n} \sim \mathcal{N}(0,1)$ appearing in Eq.\,\ref{eq:2} are independent.

\begin{figure}[ht!]
\centering\includegraphics[width=7cm]{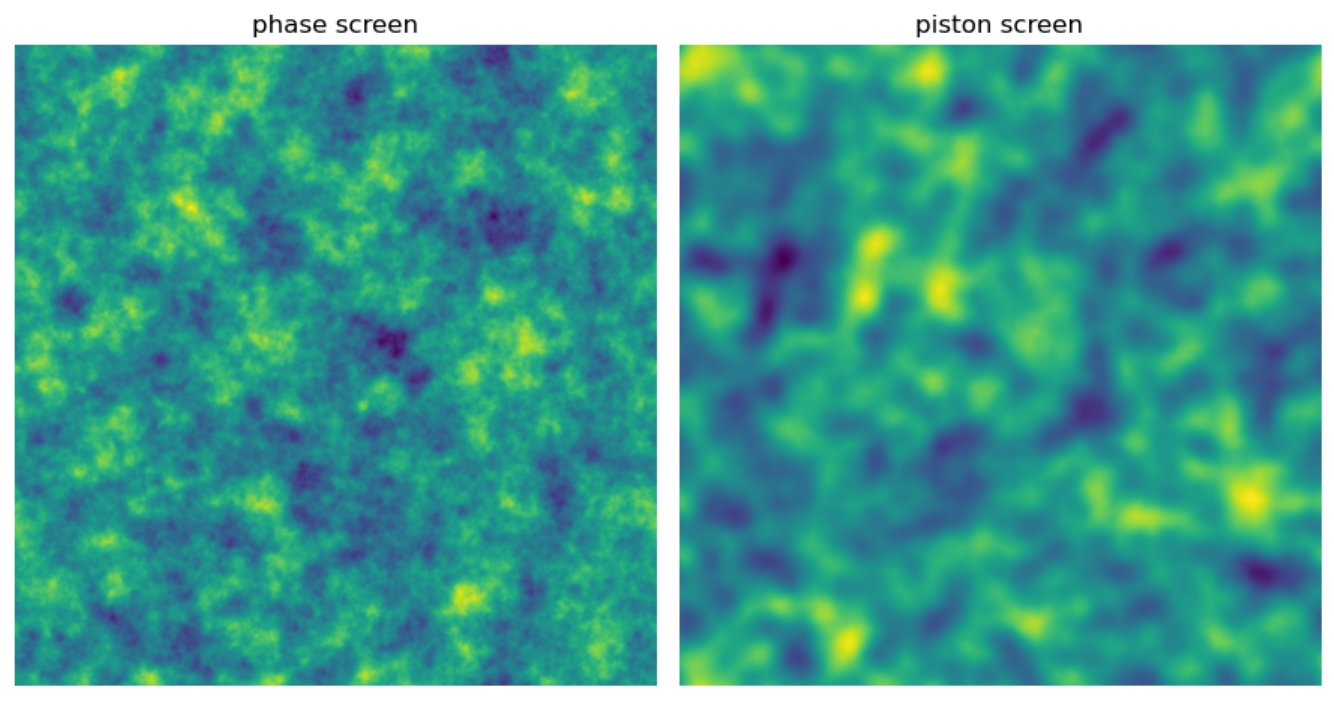}
\caption{Phase (left) and piston (right) simulations on the same scale obtained with the 2D FFT-based simulation technique.\label{fig:1}}
\end{figure}

This method can simulate not only phase fluctuations but also other optical turbulence quantities such as angle of arrival fluctuations or the piston fluctuations over a 2D spatial field, provided that the appropriate power spectrum density is used. These quantities can often be expressed in the Fourier space as a linear filtering of the phase fluctuations power spectrum density. For instance, Fig.\,\ref{fig:1} displays the results of two 2D simulations: one of piston fluctuations over a 2D field (right panel) and another of phase fluctuations (left panel). Both simulations use the same atmospheric turbulence parameters, and the impact of the telescope's pupil filtering is evident.

\subsection{Practical limitations}
\label{subsec:limitations}

The FFT-based simulation can theoretically be applied to any case, as long as the spatial frequency sampling is adjusted accordingly. However, due to limited computational resources in terms of memory usage and execution time, compromises are often made in frequency sampling. As a result, the spectral information may not be optimally transferred to the simulated quantity.

The power spectrum densities of phase fluctuations are tipically a power-law function in the form of $f^{-11/3}$. As a result, low frequencies are often under-sampled, particularly at the origin $f=0$. This represents a frequency domain of size $\delta\!f \times \delta\!f$ in the constant sampling required for the FFT algorithm, with $\delta\!f$ being the sampling pitch in the frequency domain.

To address the issue of low-frequency problems~\cite{Frehlich:00}, sub-harmonics are incorporated to compensate for the information that is not captured by the 2D FFT alone~\cite{Lane1992, Sedmak:04, Carbillet:10} and to break the FFT-induced circular symmetry. This correction is adequate for some imaging applications, even though incorporating sub-harmonics results in additional computational time penalties~\cite{Paulson:19}. Sub-harmonics can be easily implemented when dealing with phase fluctuations spectra, but they become more cumbersome for other quantities where sub-samplings do not necessarily occur at the origin of the frequency space.

Overall, under-sampling, including sub-harmonics under-sampling, because of the limited computational resources, leads to an approximation of the true power spectrum density. This sampling, as it is static, introduces systematic deviations from the desired theoretical ensemble statistics. As a consequence, for a sequence of random phase (or other quantity) screens, the same set of spatial frequencies are represented in each realisation, and the ensemble statistical properties (covariance, structure function) correspond to those sampled frequencies, not to the continuous spatial spectrum as defined by the model~\cite{Charnotskii:20}.

One way to address the issue of static spatial frequencies is to randomise the frequencies that contribute to the simulated quantity of each realisation. This approach should result in ensemble statistics that correspond to a power spectrum density covering the entire spectral continuum, as calculated from a long sequence of realisations. 

Several spatial frequency randomisation schemes have been proposed. One approach is to randomise the frequencies completely~\cite{Voitsekhovich:99}, while another is to apply a random jitter to a regular grid~\cite{Berdja2006JOptA}. It is also possible to randomly shift a regular grid~\cite{Paulson:19}, which can still be used with the 2D FFT in this case. Additionally, randomised spatial frequencies can be introduced based on the symmetry and distribution of the power spectrum density~\cite{Charnotskii:13a, Charnotskii:13b, Charnotskii:20}.

It is important to note that there is an additional consideration to take into account, in addition to the constraints that define the statistics of every spectral harmonic (refer to Section~\ref{subsec:spectrumbasedsimul}). To adhere to the energy cascade model of fully developed turbulence, each simulation realisation should \textit{ideally} incorporate all spatial frequencies.

Spectral continuity is implicitly assumed in FFT-based simulations due to the regular sampling, but this assumption may not hold true for random sampling. To demonstrate this, one could sample a single spatial frequency for each realisation, resulting in a single 2D harmonic with a random amplitude and phase. The given realisation does not represent an optical turbulence measurement, despite the ensemble statistics converging to the theoretical model through averaging across multiple realisations. Ensemble statistics alone are necessary but not sufficient to ensure accurate simulations. Spectral continuity is also required.

A random or semi-random sampling of spatial frequencies could address the statistical bias caused by the static sampling in FFT-based simulations and introduce spectral continuity. However, it may also introduce additional effects on the simulated fields due to discrepancy (refer to Section~\ref{sec:QuasiRandom}).

\section{Low discrepancy quasi-random frequency sampling}
\label{sec:QuasiRandom}

In the following, Eq.\,\ref{eq:1} is simplified to the following form:

\begin{equation}
\label{eq:3}
\chi(\vec{r}) = \frac{2}{\sqrt{2}} \sum_{f_{x}, f_{y}} R_{n}(\vec{f}) \sqrt{\delta\!f_{x} \delta\!f_{y}W_{\chi}(\vec{f})} \cos{\left( 2\pi\left(\vec{f}\cdot\vec{r} + R_{u}(\vec{f})\right) \right)},
\end{equation}
\noindent where $f_{x}$ and $f_{y}$ are the orthogonal components of the 2D spatial frequency vector $\vec{f}$.

Note that the factor of $2$ over the square root of $2$ in this equation is due to the assumption that a sampled frequency vector $\vec{f}$ does not have a counterpart vector $-\vec{f}$ as in the FFT method for instance. If a counterpart vector were present, this factor would be $1$.

This approach does not take advantage of the optimised FFT routines, but it offers the flexibility to simulate at specific areas of interest $\Vec{r}$. For example, if a simple model of piston fluctuations in an interferometric system needs to be simulated, the effects are simulated precisely where they are observed in the 2D space field. Similarly, for imaging applications, phase fluctuations can be simulated inside an aperture or a collection of apertures, with no need to simulate in the 2D space where light is masked.
Another advantage is the flexibility in sampling the spatial frequency components required to simulate the desired quantity. For example, if the power spectrum density hypothetically has a hard low-pass cutoff frequency, there is no need to use frequencies beyond this limit. It is better to focus computational effort on sampling parts of the spectrum that provide information about the statistical properties of the simulated quantity.

\subsection{Randomly jittered grid frequency sampling}

As mentioned in Section~\ref{subsec:limitations}, when sampling the spatial frequencies randomly the ensemble statistical moments will inevitably converge, through a long sequence of simulated sequences, to the theoretical model.
If the frequencies are randomly sampled across a spectral domain, a single simulated field may have an uneven concentration of frequencies. 
The assumption of a fully developed turbulence, upon which the theoretical models of optical turbulence are built, implies however a uniform representation of harmonics throughout the whole of the spectral domain.

To correct this imbalance in the spectral distribution, one solution would be to maintain a regular grid $\vec{f'}(f'_{x}, f'_{y})$ with grid cells of size $\delta\!f'_{x} \times \delta\!f'_{y}$ to which a random jitter is introduced:

\begin{equation}
\label{eq:4}
\begin{cases}
f_{x} = f'_{x} + R_{u}(f_{x}) \delta\!f'_{x}, \\
f_{y} = f'_{y} + R_{u}(f_{y}) \delta\!f'_{y}.
\end{cases}
\end{equation}

In other words, if we subdivide the spectral domain into a grid of cells of size $\delta\!f'_{x}\times \delta\!f'_{y}$, a spatial frequency $\vec{f}(f_{x}, f_{y})$ is randomly selected from within each cell. $R_{u} \sim \mathcal{U}(0,1)$ are uniformly distributed.

\begin{figure}[ht!]
\centering\includegraphics[width=12cm]{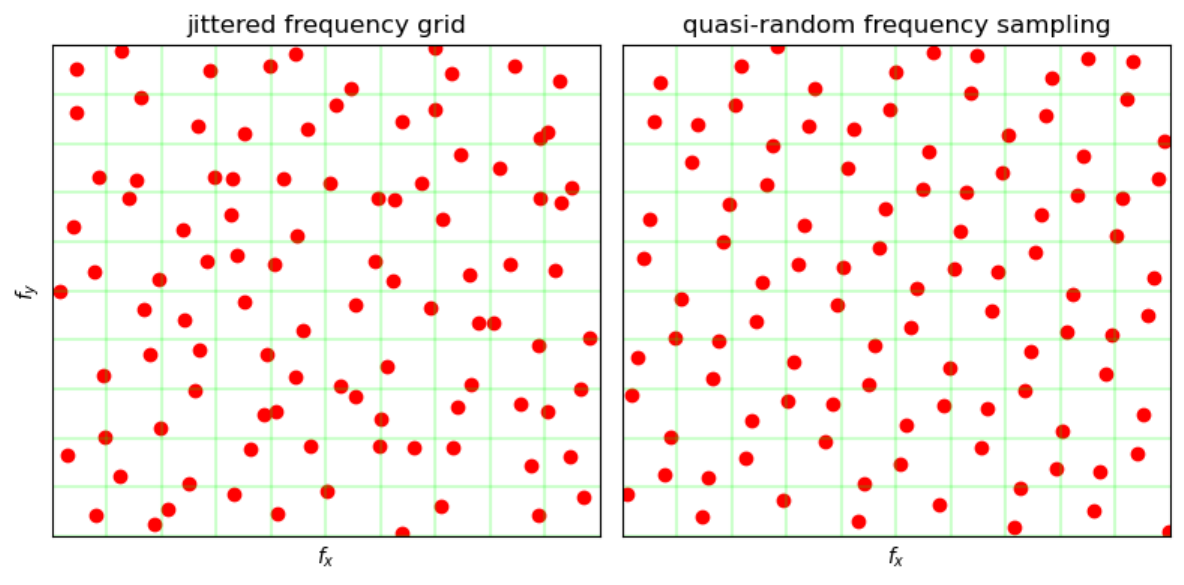}
\caption{Randomly jittered (left panel) and quasi-random (right panel) frequency sampling for one realisation. $f_{x}$ and $f_{y}$ are the axes of the 2D spatial frequency. The green lines represent a regular grid like in the FFT-based simulations. \label{fig:2}}
\end{figure}

This approach partially solves the large scale imbalance in the frequency representation, but as with taking random frequencies from the whole domain, there are many frequencies that will "clash", in the sense that they will be too close together. It is as if the same frequency is used multiple times. The left panel of Fig.\,\ref{fig:2} shows an example of a jittered regular grid where a few spatial frequencies (red dots) are so close together that they are almost indistinguishable.

When this happens, there are random gaps in the frequency coverage, but this is not a matter of concern because the amplitude of each frequency is normally distributed -- this is naturally expected to happen anyway. However, in the construction of the spectrum-based simulation approach, it is expected that the probability of each harmonic component to be bounded by the normal distribution that is assigned to it. If in the simulation of an optical turbulence quantity, some frequencies have an excessively large amplitude, even if the sequence converges statistically over many simulated fields, it is not obvious whether the individual fields have the structure that mimics the structure that can be observed in nature. Unless this is experimentally observed or theoretically jusified, we want to avoid this so-called "clashing" of frequencies.

To avoid this clustering (or clashing) of spatial frequencies, we propose instead to use low-discrepancy sequences (quasi-random numbers) to sample the spatial frequencies.

\subsection{Quasi-random (low-discrepancy) frequency sampling}

We propose the use of low-discrepancy sequences, also known as quasi-random sequences (in contrast to pseudo-random numbers) to sample spatial frequencies in optical turbulence simulations as expressed in Eq.\,\ref{eq:3}. Quasi-random sequences are fully deterministic sequences that have the property of progressively covering, as uniformly as possible, the entire domain on which they are defined. Pseudo-random sequences are designed to give the impression that a generated number has no "memory" of the numbers preceding it in the sequence. In quasi-random sequences, however, each number avoids the preceding number as much as possible. The net result is that pseudo-random sampling, including its jittered grid variant, produces clusters and gaps in the spatial domain. Quasi-random sampling, while deterministic, is not repetitive, and covers the spectral range evenly without clusters (frequency clashes) and gaps. They have exactly the desired characteristics outlined above.

To our knowledge, these sequences have mainly been used to overcome the so-called "curse of dimensionality" in high-dimensional integration~\cite{niederreiter1992}. We argue that they are a very useful tool for the numerical simulation of optical turbulence phenomena, overcoming the difficulties encountered in previous approaches.

Despite the many quasi-random sequences available in the literature and in the scientific libraries (Halton, Hammersley, Sobol), we choose to experiment with the $R_{2}$-sequence presented by Roberts~\cite{Roberts2018} for two main reasons:

First, it is the simplest and easiest to implement in code. Second, as an irrational fraction method it offers an attractive intuitive rational.

Consider a given periodic function $f_{1}(t)$ with period $T$ and a similar function $f_{2}(t)$ with a period $\alpha T$. If the functions coincide at time $t_{0}$, the next time they coincide will be at time $t_{0}+ n T = t_{0}+ m \alpha T$ where $n$ and $m$ are integers and $\alpha = n/m$. If $\alpha$ is irrational then the functions will never coincide again after $t_{0}$. When the outputs are reduced (in the modulo sense) to a bounded domain of length $T$, the output of $f_{2}(t)$ will continuously fill the interval without repetition as $t$ becomes larger. Of course, this condition is not sufficient to ensure low discrepancy, which for any such a sequence needs to be tested or demonstrated. 

The $R_{2}$-sequence can be expressed, to be consistent with the previous formulations, as follows:

\begin{equation}
\label{eq:5}
\begin{cases}
f_{x} = \delta\!f_{x} n_{x} (s_{x} - 0.5), \\
f_{y} = \delta\!f_{y} n_{y} (s_{y} - 0.5),
\end{cases}
\end{equation}

\noindent where $n_{x}$ and $n_{y}$ are natural numbers such that $\delta\!f_{x} n_{x} = 2 max(f_{x})$ and $\delta\!f_{y} n_{y} = 2 max(f_{y})$, thus defining the limits of the frequency domain.

Here, $s_{x}$ and $s_{y}$ are the 2D quasi-random numbers, between $0$ and $1$, which in the context of the $R_{2}$-sequence approach are simultaneously updated $n_{x} \times n_{y}$ times with the following additive recurrence:

\begin{equation}
\label{eq:6}
\begin{cases}
s_{x} = \left\{ s_{x}+ \frac{1}{\psi} \right\},\\
s_{y} = \left\{ s_{y}+ \frac{1}{\psi^{2}} \right\},
\end{cases}
\end{equation}

\noindent where $\{\}$ denotes taking the fractional part. $\psi = 1.324717957244746...$ is the so-called the plastic number (the positive root of the $x^{3}=x+1$ equation), an irrational number only limited by the machine's floating-point precision.

The right panel in Fig.\,\ref{fig:2} shows a realisation of a quasi-random sampling. It is worth noting that contrarily to the randomly jittered realisation in the panel above, there are no visible frequency clusters and gaps.

The initial seed for the quasi-random sequence can be set to $(s_{x}, s_{y}) = (0, 0)$ or $(s_{x}, s_{y}) = (0.5, 0.5)$, according to Roberts~\cite{Roberts2018}, but it could also be set to any random value according to our numerical experiments. 
When simulating different realisations, it is important that the quasi-random numbers are not reinitialised. One can draw an analogy with the pseudo-random numbers that can be seeded once for the whole sequence of simulations.
In fact, $s_{x}$ and $s_{y}$ should be continuously updated throughout the whole process, even when simulating a long series of fields, because on the long run, the quasi-random frequencies will cover the whole of the spectral range even so finely. It is precisely this property that allows a faster convergence to more accurate statistical quantities.

In case the power spectrum density needs to be expressed in polar coordinates, the quasi-random frequencies $(f_{x}, f_{y})$ can be transformed into polar quasi-random frequencies through a "square to disc" mapping~\cite{Shirley1997}.

\begin{figure}[ht!]
\centering\includegraphics[width=12cm]{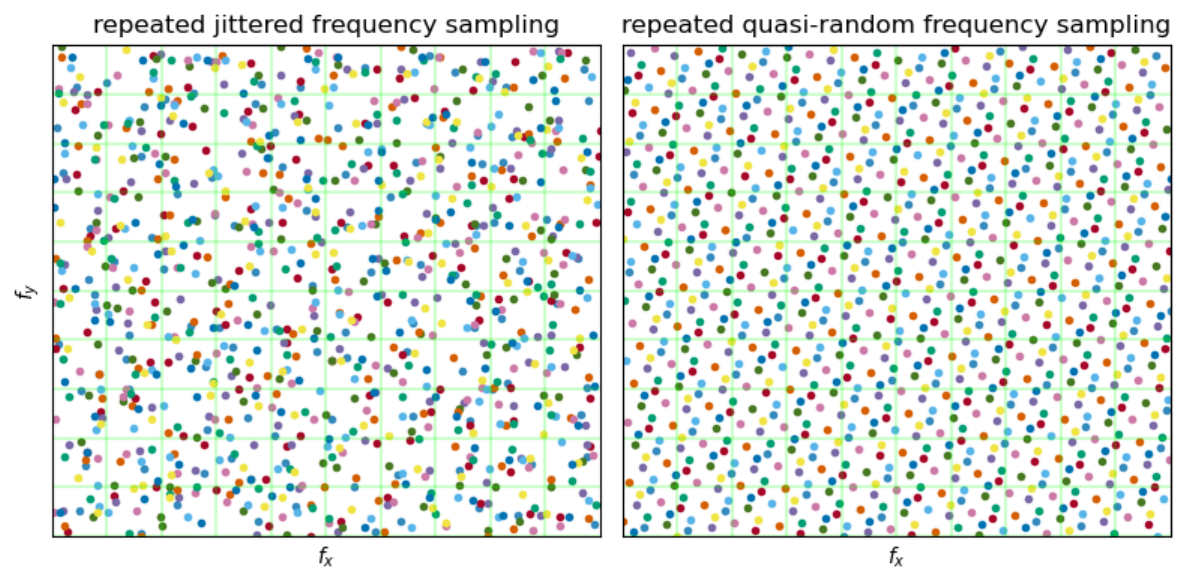}
\caption{Randomly jittered (left panel) and quasi-random (right panel) frequency sampling for several realisations. Each colour represents a different realisation. $f_{x}$ and $f_{y}$ are the axes of the 2D spatial frequency. \label{fig:3}}
\end{figure}

Fig.~\ref{fig:3} shows a comparison between, over many realisations of the simulation (represented by different colours), the jittered grid sampling (left panel) and the quasi-random sampling (right panel). The jitterd grid still has clusters and gaps in the 2D frequency coverage. The quasi-random sampling, on the other hand, in addition to being evenly distributed for each realisation (each colour separately) covers the frequency space even more finely due to its non-repeating nature, while maintaining the even distribution.

The formulation of the 2D quasi-random frequency sampling, as shown in Eq.\,\ref{eq:5}, can be simplified in most applications by taking $\delta\!f_{y} = \delta\!f_{x} = \delta\!f$ and $n_{x} = n_{y} = N$. This leaves us with two free "hyper-parameters" $\delta\!f$ and $N$.
These two parameters, under this formulation, give the flexibility to choose the spectral content of the simulation. Once $\delta\!f$ has been chosen, a value can be set for $N$, where $max(f_{x,y}) = N \delta\!f$ being the highest frequency in either axis, according to the power spectrum density and the required numerical precision of the simulation. The next section presents an example of a simulation describing this process.

\section{Application Example}
\label{sec:Application}

\subsection{Piston fluctuations model}

To demonstrate the practicality of using quasi-random sequences for optical turbulence simulations, we present an example of simulating monochromatic piston fluctuations for a long baseline stellar interferometer.

The simulation is designed to simulate a fringe-tracker~\cite{Petrov2022, Hadjara2022}, which measures the differential pistons in a specific wavelength band. This allows for near-instantaneous corrections to be made in another band where the "science object" is observed. The discussion at hand does not cover the simulation details of the fringe tracker, which include polychromatic dispersion, beam-combination, angular dispersion, anisoplanatism, measurement errors, and fringe wander reconstruction.

The benefit of this option is its simplicity. If we were simulating imaging through a telescope or a Michelson interferometer, we would need to discuss minor technical details such as aperture sampling and vectorisation. However, since this is not the case, we can exclude these details from our discussion.

In this example, there is a single turbulent layer that moves at a constant velocity over the telescopes. According to the Taylor frozen turbulence hypothesis, the temporal statistical properties are equivalent to the spatial properties along the wind orientation. In other words, a time interval $\tau$ is equivalent to a distance of $x = \tau v_{w}$ where $v_{w}$ is the wind velocity. $x$ is measured along the wind's direction as represented in Fig.\,\ref{fig:4}. Although turbulence cannot remain frozen during a long time, for practical purposes, we are only interested in the instrument's response to differential piston fluctuations (or equivalently, Optical Length Difference variations) over times in the range of a few tens of milliseconds. Therefore, the Taylor hypothesis is appropriate for this simulation.

A simplified model of the spatial power spectrum density of piston fluctuations based on the von\,K\'{a}rm\'{a}n phase fluctuations model~\cite{Lukin2021} is given by:

\begin{equation}
\label{eq:7}
W_{\overline{\varphi}}(\vec{f}) \approx 0.02289\ r_{0}^{-\frac{5}{3}} \left( f^{2} +L_{0}^{-2} \right)^{-\frac{11}{6}} 4\left( \frac{D J_{1}(\pi D f)- d J_{1}(\pi d f) )  }{\pi (D^{2}-d^{2}) f} \right)^{2},
\end{equation}

\noindent where $r_{0} = r_{0}(\lambda)$ is the Fried parameter at wavelength $\lambda$, $L_{0}$ is the outer scale of optical turbulence, $D$ is the telescope's diameter, and $d$ is the telescope's central obstruction diameter. $J_{1}(...)$ is the first kind Bessel function of order 1.

The von\,K\'{a}rm\'{a}n power spectrum density of phase fluctuations is represented by the first term in Eq.\,\ref{eq:7} with the exponent $-11/6$. The equation's second term, which has an exponent of $2$, represents the aperture filtering. In this case, the aperture is modelled as a circular aperture with a diameter of $D$ and a circular central obstruction with a diameter of $d$.

Applying the procedure outlined in Eq.\,\ref{eq:3} and Eq.\,\ref{eq:5} is straightforward with this model. However, the hyper-parameters $N = n_{x} = n_{y}$ and $\delta\!f = \delta\!f_{x,y}$ require a choice to be made.

\subsection{Hyper-parameters}

The hyper-parameters $\delta\!f$ and $N$ are selected based on the properties of the power spectrum density $W_{\overline{\varphi}}(\vec{f})$ that need to be expressed in the simulation.

Long-baseline interferometry is sensitive to piston fluctuations over distances that are of the order of the outer scale $L_{0}$. To ensure consistent sampling of low frequencies in the von K\'{a}rm\'{a}n's model during simulation, a reasonable value for $\delta\!f$ is $\delta\!f = 0.25/L_{0}$. 

To determine $N$, such that $2 max(f_{x,y}) = N \delta\!f$, it is possible to follow the following steps: Since in this case the power spectrum density is isotropic ($W_{\overline{\varphi}}(\vec{f}) = W_{\overline{\varphi}}(f)$), it is possible to calculate (from the model) the quantity $g(x) = \int_{0}^{x} df' f' W_{\overline{\varphi}}(f')$. Due to the finite variance of the von K\'{a}rm\'{a}n model and to the aperture filtering of the telescope, the function $g(x)$ saturates and converges to a finite variance $\sigma_{\overline{\varphi}}^{2}$ at $f' \rightarrow \infty$. A stopping criterion can be defined such that $g(x)-g(x-\delta\!f) < \epsilon$, where $\epsilon << \sigma_{\overline{\varphi}}^{2}$. 

There are other possible criteria that depend on the required precision of the simulated fluctuations. This provides additional flexibility to the simulation approach, allowing for a choice between precision and speed.

\subsection{Numerical Implementation}

The implementation of the simulation algorithm is straightforward given a numerical function \verb|sqrt_w| that calculates the absolute square root of the power spectrum density $W_{\overline{\varphi}}(\vec{f})$.

\begin{figure}[htbp]
\centering\includegraphics[width=12cm]{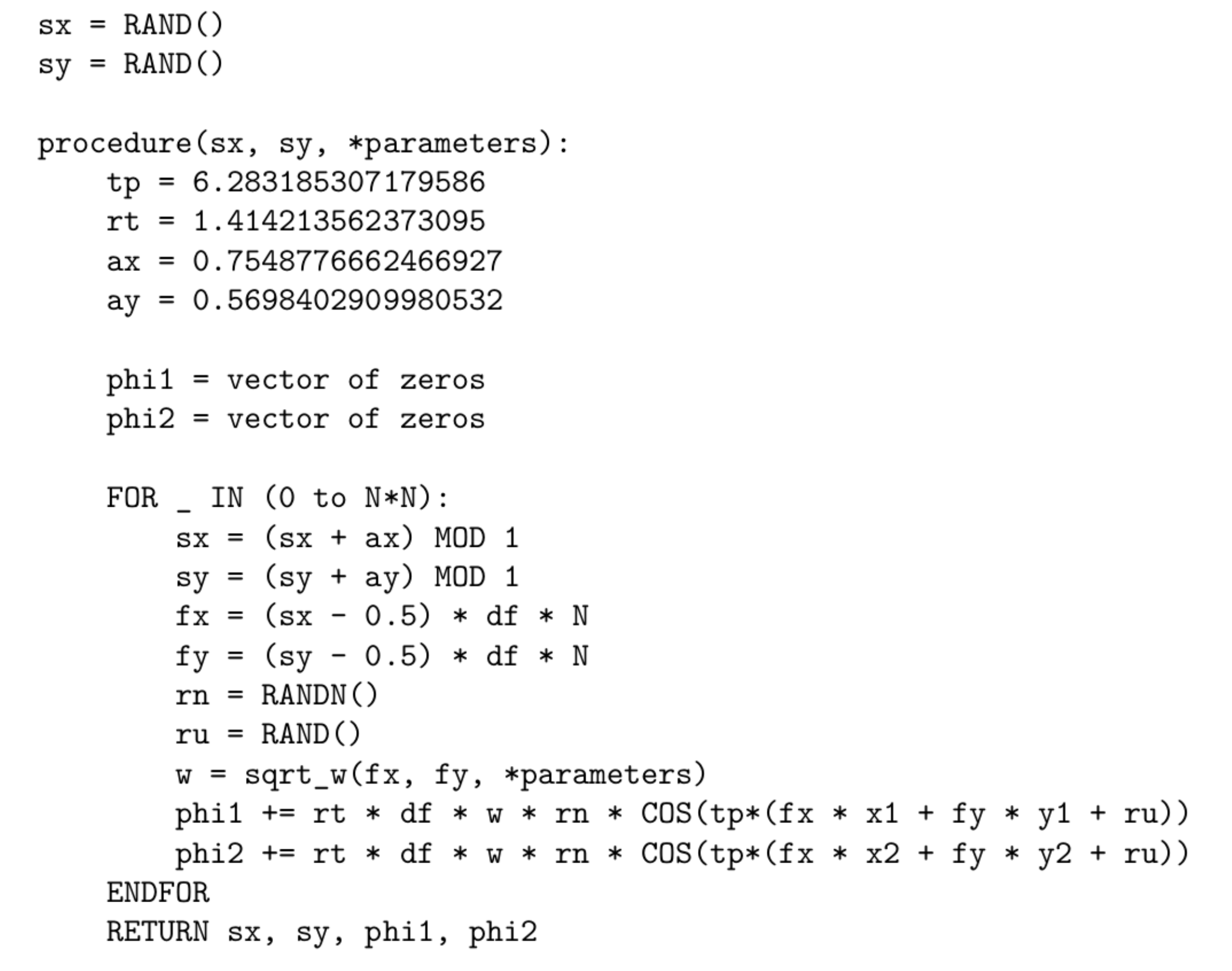}
\caption{A pseudo-code of the simulation procedure using quasi-random numbers. \label{fig:4b}}
\end{figure}

Fig.\,\ref{fig:4b} is a pseudo-code of the simulation algorithm. It can easily be implemented in any programming language.

The power spectrum density function receives optical turbulence parameters such as $r_{0}(\lambda)$ and $L_{0}$ as inputs. The vectors \verb|xi| and \verb|yi| ($i = 1,2,...$) represent the 2D position components where pistons are simulated.

The positions sampled depend on the simulation's needs. In this case, they are selected to achieve millisecond time sampling. Unlike in the FFT-based method, this sampling is not bound to the sampling in the Fourier space, providing greater modelling flexibility.
It is evident from this pseudo-code that the seeds \verb|sx| and \verb|sy| are global variables that are updated by the simulation process. This guarantees that the entire spectral range, delimited here by \verb|df| and \verb|N| is evenly sampled if the simulated pistons are to be used for statistical characterisation.

\begin{figure}[htbp]
\centering\includegraphics[width=12cm]{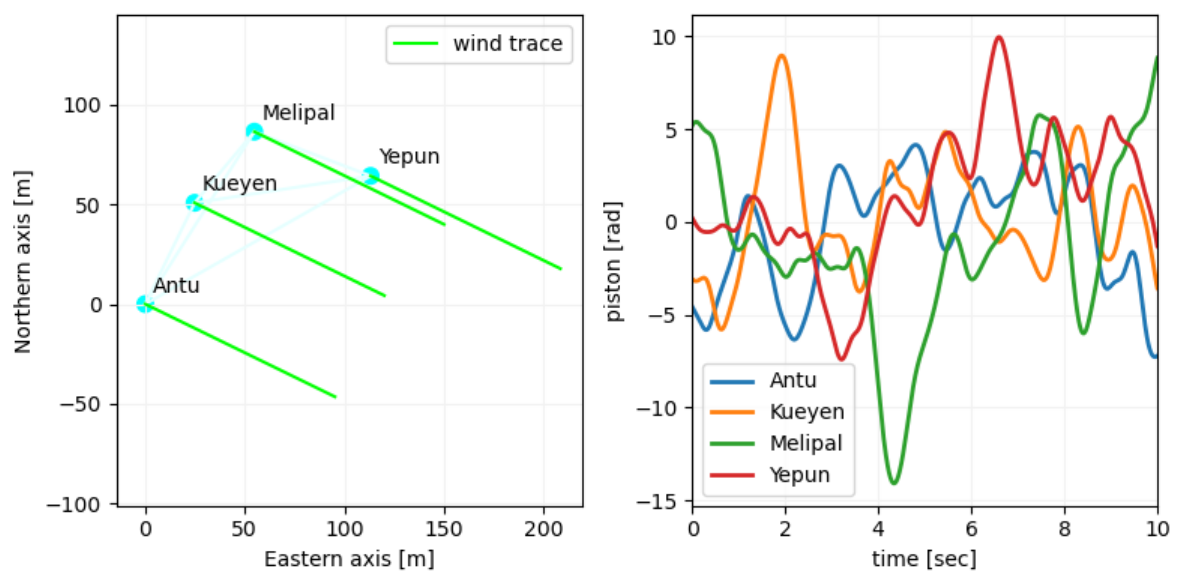}
\caption{A representation of the overall geometry for the simulation example (left panel). The four blue circles are the four UT telescopes of the VLTI as seen from above. The green lines represent the location of optical turbulence as "seen" by the telescope through time. Their direction is the wind direction. The light lines between the telescopes are the six interferometric baselines. On the right panel a single realisation of the simulation generates four simultaneous time series of piston fluctuations for each telescope. \label{fig:4}}
\end{figure}

The authors have implemented the code for these results in Python, and accelerated it with the numba JIT compiler. The codes are available upon request.

For the statistical tests presented in Section~\ref{subsec:stattest}, the parameters used for the simulation correspond to the VLTI telescopes as shown in Fig.\,\ref{fig:4} in the left panel.
VLTI is the interferometric mode of the European Southern Observatory's (ESO) Very Large Telescope (VLT) located at the Paranal observatory in Chile. The VLTI combines, in the infrared wavelength ranges, 4 large VLT telescopes called UT (Unit Telescopes), which are fixed and have a diameter of $D$ = 8m and $d$=1.1m central obstruction each, for a maximum baseline of 130m. The VLTI also has an auxiliary array of 4 ATs (Auxiliary Telescopes) with a diameter of $1.8m$ that can be moved to a large number of stations on the Paranal platform with a maximum baseline of 200m. In the simulation we are only concerned with piston fluctuations at the VLT telescopes.

The right panel in Fig.\,\ref{fig:4} shows a single realisation of the simultaneous simulated piston fluctuations at the 4 telescopes of the VLTI.
The smoothness of the curves is due to the aperture filtering of the telescopes and the wind velocity as the turbulence layer moves above them. For the simulation in Fig.\,\ref{fig:4}, as well as the one leading to the results in Fig.\,\ref{fig:5}, wind speed is $10.2m/s$ and the outer scale $L_{0}$=22m. The hyper-parameters $\delta f = 0.25/22 m^{-1}$ and $f_{max}$ = 0.25$m^{-1}$, hence $N$=44. In other conditions and with other optical turbulence quantities like phase fluctuations, these parameters would be different.

\subsection{Statistical test}
\label{subsec:stattest}

The simulation's geometry produces a single time series for every telescope, enabling a straightforward statistical test. For each pair of telescopes $i$ and $j$, the structure function of the differential piston fluctuations $D_{\Delta\overline{\varphi}}^{i, j}(\tau)$ is given by $\left< \left| \Delta\overline{\varphi}_{i, j}(t) \Delta\overline{\varphi}_{i, j}(t+\tau) \right|^{2} \right>_{t}$. $<...>_{t}$ represents an average over all the available values $t$ on the same simulation realisation.
$\Delta\overline{\varphi}_{i, j}(t)$ are the differential piston fluctuations between telescopes $i$ and $j$, given by $\overline{\varphi}_{i}(t) - \overline{\varphi}_{j}(t)$.

The length of the simulated piston fluctuations time series is finite. Therefore, the statistical error, which depends on the number of points used to for every separation $\tau$ and the coherence time, increases with $\tau$. If there are $M$ points separated by $\delta \tau$, $D_{\Delta\overline{\varphi}}^{i, j}(\delta \tau)$ is calculated using $M-1$ data pairs, resulting in lower statistical error, while and $D_{\Delta\overline{\varphi}}^{i, j}((M-1)\delta \tau)$ is calculated only once, resulting in a higher statistical error. If the one-realisation structure function converges to the theoretical structure function for an infinite length of the simulated time series, then the statistics can considered "ergodic". This can be assured by using the spectral continuity constraint with adequate sampling.

Structure functions from a large number of independent realisations are averaged out. If the simulation statistics are good, the averaged structure function will converge to the theoretical structure function. However, if the simulation statistics are biased due to inadequate sampling, the averaged structure function will converge to a function that differs from the theoretical structure function.

In Fig.\,\ref{eq:5}, the average structure function of the Optical Path Difference (OPD), $D_{OPD}^{1, 2}(\tau)$, defined as 
$D_{OPD}^{i, j}(\tau) = \left( \frac{\lambda}{2\pi} \right)^{2} D_{\Delta\overline{\varphi}}^{i, j}(\tau)$, is compared to the expected theoretical result. $\lambda$ in the wavelength of the light.
The average structure function has been obtained from 2000 independent realisations.

\begin{figure}[htbp]
\centering\includegraphics[width=12cm]{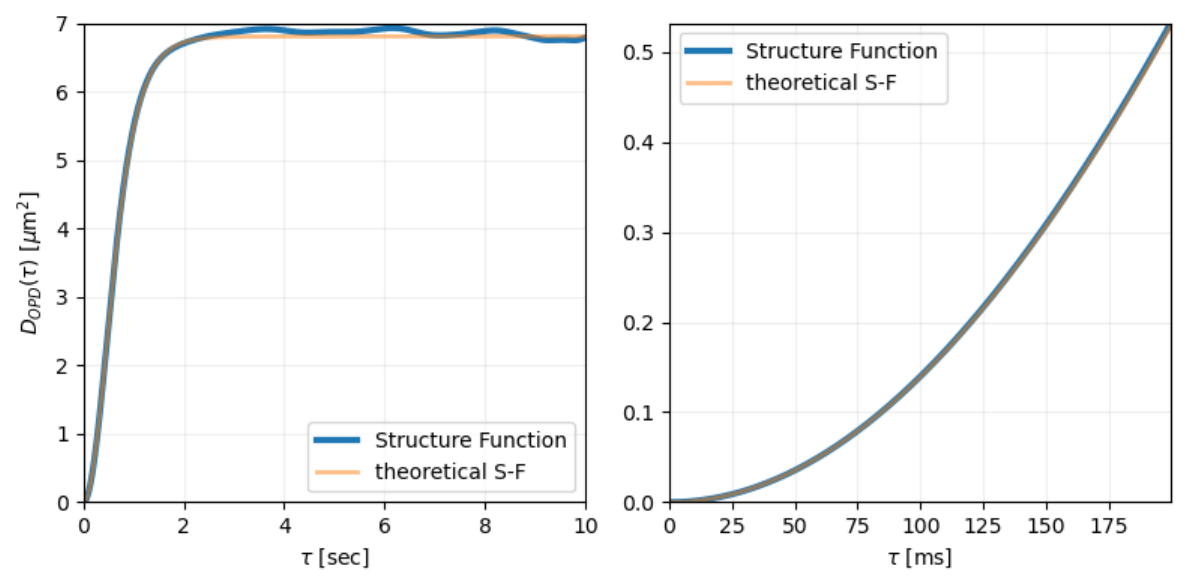}
\caption{Average structure function of the Optical Path Difference (OPD) of one of the six baselines (telescopes 1 and 2, separated by 56.5m) compared to the corresponding theoretical structure function. The right panel is a detail of the same result. With 2000 independent realisations, statistical errors are visible at large intervals. With more realisations, the curves become indistinguishable. \label{fig:5}}
\end{figure}

The structure function resulting from our simulation almost perfectly matches the theoretical structure function. For large time delays $\tau$, there are still some statistical errors (left panel of Fig.\,\ref{fig:5}). These errors are inevitable but they can be reduced with more realisations. 
At smaller $\tau$ values, the structure function is more precise (right panel in Fig.\,\ref{fig:5}).

The primary objective of the simulation is to generate data that closely resemble the expected measurements. The structure function confirms that the information from the power spectrum density is accurately translated into the simulated quantities. As discussed in Section~\ref{subsec:spectrumbasedsimul}, the structure function, especially when calculated over a large number of realisation, is necessary to assess the validity of the the simulation. However, the simulation's validity is not solely determined by the structure function. It is completed by the physics-based constraint of spectral continuity.
In this case, the model's properties are accurately captured.

\section{Conclusion}
\label{sec:conclusion}

We have presented an approach to simulate the observable fluctuations of optical turbulence using a quasi-random sampling of spatial frequencies within the framework of spectrum-based simulations.

We have also discussed how this approach allows us to overcome some of the difficulties encountered with the more common FFT-based simulations. It could be argued that it is unfortunate that we do not take advantage of the optimisations offered by the FFT algorithm. However, depending on the situation, this new approach not only does not suffer from the same limitations as the FFT-based simulations do, but it also offers some advantages.

Taking the example of imaging the output of a Michelson interferometer, the FFT-based simulation method requires generating matrices that are large enough to contain the baseline between the telescopes and sufficiently sample each telescope. The phase fluctuations used for the image are only a small fraction of the generated phase screen. The proposed quasi-random frequencies approach generates phase fluctuations only where they are useful, at the telescopes' apertures. Not using the FFT approach can be computationally advantageous in this case.

We do not expect this approach to entirely existing methods. However, when applicable, it offers clear advantages, provided that one knows how to adjust the hyper-parameters for a given task. We have mentioned the possibility of mapping the "square" quasi-random sampling to a polar quasi-random sampling for special cases of expressing the power spectrum density. Systematising hyper-parameters with a mapping that takes into account the properties of the power spectrum density may be possible, but it is not necessary at this point. Fixing these hyper-parameters on a case-by-case basis allows for more flexibility.

The simulation approach's validity has been demonstrated through differential piston fluctuations. This method can be applied to other optical turbulence measurements, such as phase or angle-of-arrival fluctuations, provided that the power spectrum density of the measurement is known, either theoretically or empirically.

\begin{backmatter}
\bmsection{Funding}
Do later.
\bmsection{Acknowledgments}
Do later.
\bmsection{Disclosures}
The authors declare no conflicts of interest.
\bmsection{Data availability} Data underlying the results presented in this paper are not publicly available at this time but may be obtained from the authors upon reasonable request.
\end{backmatter}


\bibliography{bibliography}

\end{document}